\documentclass[a4paper,11pt]{article}
\usepackage{geometry}
\usepackage{a4wide}
\usepackage{amsmath}
\usepackage[normalem]{ulem}
\usepackage{amssymb}

\usepackage{cite}
\usepackage{appendix}
\usepackage{amsfonts}
\usepackage{amsthm}
\usepackage{euscript}
\usepackage{xcolor}
\usepackage{braket}
\newcommand{\be}{\begin{equation}}
\newcommand{\ee}{\end{equation}}

\newcommand{\Rmnum}[1]{\expandafter\@slowromancap\romannumeral #1@}
\newcommand{\bea}{\begin{eqnarray}}
\newcommand{\eea}{\end{eqnarray}}

\numberwithin{equation}{section}

\begin{document}

\title{\bf A note on the total action of $4D$ Gauss-Bonnet theory}

\author{\textbf{Subhash Mahapatra}\thanks{mahapatrasub@nitrkl.ac.in}
 \\\\
 \textit{{\small Department of Physics and Astronomy, National Institute of Technology Rourkela, Rourkela - 769008, India}}
}
\date{}


\maketitle
\abstract{Recently, a novel four-dimensional Gauss-Bonnet theory has been suggested as a limiting case of the original $D$-dimensional theory with singular Gauss-Bonnet coupling constant $\alpha\rightarrow\alpha/(D-4)$. The theory is proposed at the level of field equations. Here we analyse this theory at the level of action. We find that the on-shell action and surface terms split into parts, one of which does not scale like $(D-4)$. The limiting $D\rightarrow4$ procedure, therefore, gives unphysical divergences in the on-shell action and surface terms in four dimensions. We further highlight various issues related to the computation of counterterms in this theory.}

\section{Introduction}
Higher-order curvature terms are expected to play a central role in quantum gravity. It is generally expected that the low-energy expansion of the quantum gravity, such as string theory,  will provide an effective Lagrangian containing higher-order curvature terms \cite{Gross:1986mw}. Finding and analysing the solutions of higher curvature Lagrangian are therefore of great physical interest.
\\
\\
One of the most studied higher curvature Lagrangian is the Gauss-Bonnet combination
\begin{eqnarray}
\mathcal{L}_{GB} = R_{\mu\nu\rho\sigma}R^{\mu\nu\rho\sigma}-4R_{\mu\nu}R^{\mu\nu}+R^2\,.
\label{actionGB}
\end{eqnarray}
With $\mathcal{L}_{GB}$, the Einstein equations of motion still remain second order in metric and it provides the simplest non-trivial modification of general relativity.
The Einstein-Gauss-Bonnet gravity action \cite{Boulware:1985wk,Deser:2002jk,Deser:2003up,Cai:2001dz}
\begin{eqnarray}
S_{EGB} =  -\frac{1}{16 \pi G_D} \int \mathrm{d^D}x \ \sqrt{-g}  \ \left[R + \frac{(D-1)(D-2)}{L^2} + \alpha \left(R_{\mu\nu\rho\sigma}R^{\mu\nu\rho\sigma}-4R_{\mu\nu}R^{\mu\nu}+R^2 \right) \right]\,. \nonumber\\
\label{action}
\end{eqnarray}
leads to the following field equation
\begin{eqnarray}
R_{\mu\nu}-\frac{1}{2}g_{\mu\nu} R - g_{\mu\nu} \frac{(D-1)(D-2)}{2L^2} + \alpha \mathcal{H}_{\mu\nu} \,.
\label{EOM}
\end{eqnarray}
and admits a consistent and non-trivial solutions for $D\geq 5$
\begin{eqnarray}
ds^2 = -f(r)dt^2 + \frac{dr^2}{f(r)} + r^2 d\Omega_{D-2}^2,  \ \ \ \  D\geq 5\,.
\label{solg}
\end{eqnarray}
where $d\Omega_{D-2}^2$ is the unit metric of the $(D-2)$-dimensional sphere. In eq.~(\ref{EOM}), the Gauss-Bonnet contribution to field equations reads
\begin{eqnarray}
& & \mathcal{H}_{\mu\nu} =  2\left(R R_{\mu\nu}-2R_{\mu\alpha\nu\beta}R^{\alpha\beta}+R_{\mu\alpha\beta\gamma}R_{\nu}^{\ \ \alpha\beta\gamma} -2 R_{\mu\alpha} R_{\nu}^{\ \alpha} \right) \nonumber\\
& & -\frac{1}{2}g_{\mu\nu} \left(R_{\alpha\beta\rho\sigma}R^{\alpha\beta\rho\sigma}-4R_{\alpha\beta}R^{\alpha\beta}+R^2 \right) \,.
\label{GBtensor}
\end{eqnarray}
Importantly, the Gauss-Bonnet term reduces to the Euler number (or to a total derivative term) in four dimensions and therefore does not contribute to the field equations. In particular, $\mathcal{H}_{\mu\nu}$ vanishes identically in $D=4$. Therefore, it came as a big surprise when a four-dimensional Einstein-Gauss-Bonnet theory was constructed in \cite{Glavan:2019inb}. The authors of \cite{Glavan:2019inb} suggested that (i) by  rescaling the Gauss-Bonnet coupling parameter $\alpha\rightarrow \tilde{\alpha}/(D-4)$, and then (ii) taking the limit $D\rightarrow 4$, a non-trivial black hole solution
\begin{eqnarray}
& & ds^2 = -f(r)dt^2 + \frac{dr^2}{f(r)} + r^2 d\Omega^{2} \nonumber\\
& & f(r)= 1 + \frac{r^2}{2\tilde{\alpha}} \left[1\pm \sqrt{1+ 4\tilde{\alpha} \left(\frac{2M}{r^3}-\frac{1}{L^2} \right)}   \right]\,.
\label{solg4}
\end{eqnarray}
in four dimensions can be obtained as a limiting case of the $D$-dimensional theory. Interestingly, this novel four-dimensional Gauss-Bonnet theory was suggested to bypasses the Lovelock's theorem \cite{Lovelock:1971yv,Lovelock:1972vz,Lanczos:1938sf} and, therefore, has created a lot of excitement in the gravitational community. \\

The essential idea behind the work of \cite{Glavan:2019inb} was the observation that the Gauss-Bonnet tensors $\mathcal{H}_{\mu\nu}$ scale like $(D-4)$ in $D$-dimensions and this $(D-4)$ factor can be cancelled consistently in the field equations by modifying the coupling parameter $\alpha \rightarrow \tilde{\alpha}/(D-4)$. At the equation of motion level, the four-dimensional Gauss-Bonnet theory was suggested as a limiting case of the original $D$-dimensional theory
\begin{eqnarray}
\lim_{D\rightarrow 4}\left[ R_{\mu\nu}-\frac{1}{2}g_{\mu\nu} R - g_{\mu\nu} \frac{(D-1)(D-2)}{2L^2} + \frac{\tilde{\alpha}}{D-4} \mathcal{H}_{\mu\nu} \right] =0 \,.
\label{GLD4eq}
\end{eqnarray}

The suggested four-dimensional gravity has already intrigued a large amount of research work in applications, see for a necessarily biased selection \cite{Konoplya:2020bxa,Guo:2020zmf,Fernandes:2020rpa,Wei:2020ght,Wei:2020poh,Casalino:2020kbt,Kumar:2020owy,Hegde:2020xlv,Ghosh:2020syx,Konoplya:2020juj,1791278,Aragon:2020qdc,HosseiniMansoori:2020yfj,
Zhang:2020sjh,EslamPanah:2020hoj,Jin:2020emq,Nojiri:2020tph,Mishra:2020gce,Churilova:2020aca,Roy:2020dyy,Kobayashi:2020wqy,Heydari-Fard:2020sib,Li:2020tlo,Kumar:2020xvu,Liu:2020vkh,Zhang:2020qam,
NaveenaKumara:2020rmi,Ying:2020bch}. Recently, many works addressing various ambiguities, even at the level of field equations, when applying the method of \cite{Glavan:2019inb} have also started appearing, in particular, see \cite{Gurses:2020ofy,Ai:2020peo,Fernandes:2020nbq,Hennigar:2020lsl,Aoki:2020lig,Alkac:2020zhg,Casalino:2020pyv,Bonifacio:2020vbk,Arrechea:2020evj,Liu:2020evp,Lu:2020iav,Shu:2020cjw}.\\

Since the Gauss-Bonnet term also contributes to local dynamics in the dimensional limiting procedure of \cite{Glavan:2019inb}, thereby dethroning the Einstein gravity as a unique covariant theory of gravity in four spacetime dimensions, greater and rigorous scrutiny of this type of limiting procedure in various physical scenarios is therefore essential to validate the claims of proposed four-dimensional gravity theory. Undeniably, there are many reasons to believe that this limiting procedure might not be sensible. For instance, since the spacetime tensor indices behave discretely and depend on the spacetime dimensions, there is no continuous way to take the suggested limit $D\rightarrow 4$ in the higher dimensional action and equation of motion. By taking the limit $D\rightarrow4$, one is not only breaking the Riemannian geometric foundations of gravity but also compromises the democratic nature of the higher dimensional spacetime coordinates.  Perhaps, one of the strongest arguments against the insensible nature of the $D\rightarrow 4$ limiting procedure is that it goes against the very essence of Lovelock's theorem \cite{Lovelock:1971yv,Lovelock:1972vz}. In particular, the gravity action with an integrand which is quadratic in the curvature component, from which the entire set of Riemannian geometry in four dimensions can be extracted, can be a linear combination of only two terms ($R^2$ and $R_{\mu\nu}R^{\mu\nu}$), \textit{i.e.} all other curvature invariant terms (such as $R_{\mu\nu\rho\lambda}R^{\mu\nu\rho\lambda}$) are redundant in the formation of field equations in four dimensions \cite{Lanczos:1938sf}. However, the same is not true in higher dimensions where other quadratic curvature invariants can exist. The Lovelock's theorem, therefore, does directly imply the impossibility of obtaining a consistent four-dimensional quadratic curvature action from the naive $D\rightarrow4$ limiting procedure.\\

It is important to emphasize that the novel four dimensional theory was suggested at the level of field equation. In particular, the defining equation of the four dimensional Gauss-Bonnet theory was suggested as a $D\rightarrow4$ limiting case of the original $D$-dimensional field equation.  Moreover, in this suggested theory, one further has to assume or demand some symmetries at the level of $D$-dimensional solution, such that (or rather hope that) the structure of the rest $(D-4)$ fiducial dimensions does not appear in the four dimensional solution. This could be the case, for example, for conformally flat geometries, including the FLRW solution or maximally symmetric solutions. However, in general, this is not the case. The prescription of \cite{Glavan:2019inb}, even if correct, undoubtedly corresponds to a highly constrained gravity set-up. Since the Lovelock's theorem does not directly apply to such a constrained setup, the work of \cite{Glavan:2019inb} therefore does not suggest the breakdown of Lovelock's theorem. Indeed, as shown in \cite{Gurses:2020ofy}, there is no covariant Gauss-Bonnet tensor in four-dimensions as assured by the Lovelock theorem. \\

Even though the four dimensional Gauss-Bonnet theory was formulated at the level of field equations, nonetheless, it is instructive and important to probe different aspects of this theory, particularly to those which are not restricted to the field equation alone. This will not only add other important directions in the discussion of four dimensional Gauss-Bonnet theory but also help to find further flaws or strengths of this theory at a more fundamental level. In this work, we investigate one such aspect of this theory. In particular, here we explore and scrutinise the four-dimensional Gauss-Bonnet theory from the action point of view.\\

Our main aim here is to analyse how the total action, consisting of various surface terms and counterterms in addition to the Einstein-Gauss-Bonnet action, with singular coupling $\alpha$ behaves in $D\rightarrow4$ limit. The action analysis is essential to understand whether the theory is fundamentally in good shape or not. As is well known, the action (\ref{action}) has to be supplemented by the surface terms to have a well defined variational problem. These surface terms although do not modify the field equation, however, they are an integral part of the gravity action. In other words, the Einstein-Gauss-Bonnet action alone is not sufficient to be considered as a well-defined theory and it need to be endowed by the various surface terms -- otherwise, there is no well defined variation problem to the action itself, which ultimately makes the theory ill-defined -- and these terms have to be investigated thoroughly. Similarly, counterterms are needed in the total action to make it IR finite. In particular, both on-shell action and surface terms suffer from infinities as the boundary is taken to infinity and  these infinities can be removed by adding local counterterms in the action. \\

Since the surface and counterterms live in one lower dimension, \textit{i.e.} at the boundary in $(D-1)$ dimensions, therefore one might expect that unlike various tensors (constructed from the $D$-dimensional metric) appearing in Einstein's field equation (\ref{EOM}), not all the surface and counterterms would come with a multiplicative $(D-4)$  factor. If that is the case then the singular coupling constant $\alpha$ will make the whole action divergent in the limit as $D\rightarrow4$. Our analysis suggests that this is indeed the case. In particular, both on-shell and surface terms diverge in the limit $D\rightarrow4$. We further highlight various issues related to the computation of counterterms for the four-dimensional Einstein-Gauss-Bonnet theory.\\

Another important point that we like to emphasize is that in this work we are not explicitly presenting the evidence for the flawed and unphysical nature of the four dimensional field equations (eq.~(\ref{GLD4eq})). In fact, we will not even use the limiting four dimensional field equations and their solutions. Instead, we will take the Gauss-Bonnet action and the corresponding solution in $D$-dimensions to show that there are fundamental flaws in the $D\rightarrow4$ limiting procedure. The prescription and methodology of our work, therefore, should be contrasted from the prescription of \cite{Glavan:2019inb}.

\section{Total action of four-dimensional Einstein-Gauss-Bonnet gravity}
Before separately analysing each term of the total Einstein-Gauss-Bonnet gravity action, it is useful to first note down the expression of $\mathcal{L}_{GB}$ and Ricci scalar $R$ in $D$-dimensions. For the metric (\ref{solg}), we have \footnote{Here the Riemann and Ricci tensor sign conventions $\left[\nabla_{\mu}, \nabla_{\nu} \right] V_{\lambda} = V_{\sigma} R^{\sigma}_{ \ \lambda\mu\nu}$ and $R_{\mu\nu}=R^{\rho}_{ \ \mu \rho \nu}$ are used.}
\begin{eqnarray}
\mathcal{L}_{GB} = \frac{2(D-2)(D-3)}{r^2}\left[f f' - f' \right]' + \frac{4(D-2)(D-3)(D-4)}{r^3} \times \nonumber\\ \left[f'(f - 1) + \frac{(D-5)}{4 r}(f - 1)^2 \right]  \,.
\label{LGBD}
\end{eqnarray}
\begin{eqnarray}
R= \frac{(D-2)(D-3)}{r^2}(1-f)-\frac{2(D-2)f'}{r}-f'' \,.
\label{RicciD}
\end{eqnarray}
here, and in the subsequent subsections, we write results explicitly in terms of $D$ and $f$. This will help to analyse the limit $D\rightarrow4$ in a clear and straightforward way, as the function $f$ is well defined in this limit. For simplicity, we derive results by assuming the black hole background. The analysis can be straightforwardly generalised to pure AdS spaces.

\subsection{The on-shell action}
The first indication that the Einstein-Gauss-Bonnet gravity action (\ref{action}) is not well defined in the limit $D\rightarrow 4$ can be seen by evaluating the on-shell action,
\begin{eqnarray*}
S_{EGB}^{\text{on-shell}} =  -\frac{1}{16 \pi G_D} \int \mathrm{d^D}x \ \sqrt{-g}  \ \left[\frac{-2(D-1)}{L^2} + \frac{2\alpha}{D-2} \mathcal{L}_{GB} \right] \,,
\end{eqnarray*}
\begin{eqnarray}
& & S_{EGB}^{\text{on-shell}}  = -\frac{1}{16 \pi G_D} \int \mathrm{d^D}x \ r^{D-2} \biggl[\frac{-2(D-1)}{L^2} + \frac{8\alpha (D-3)(D-4)}{r^3} \times \nonumber\\
& & \hspace{2cm}  \left(f'(f - 1) + \frac{(D-5)}{4 r}(f - 1)^2 \right)
  +  4\alpha (D-3)\frac{\left(f f' - f' \right)'}{r^2}  \biggr] \,.
\label{onshellaction1}
\end{eqnarray}
If we let $\alpha = \tilde{\alpha}/(D-4)$ and take the limit $D\rightarrow 4$, then the second term in eq.~(\ref{onshellaction1}) gives finite contribution and is well defined. On the other hand, the last term is although a total derivative term (hence does not contribute to the field equations), however, gives infinite contribution. Another way to see that the on-shell Einstein-Gauss-Bonnet gravity action is not well defined in the limit $D\rightarrow 4$ is by noticing that it can also be rewritten as \footnote{This form of the on-shell action is obtained by first taking the trace of the Einstein equation and then solving for $\mathcal{L}_{GB}$. Substituting the obtained $\mathcal{L}_{GB}$ into the action, we get the desired on-shell action (\ref{onshellaction2}).}
\begin{eqnarray}
S_{EGB}^{\text{on-shell}} =  -\frac{1}{16 \pi G_D} \int \mathrm{d^D}x \ \sqrt{-g}  \ \left[\frac{-2 R}{(D-4)L^2} - \frac{4(D-1)(D-2)}{(D-4)L^2} \right] \,.
\label{onshellaction2}
\end{eqnarray}
here we have substituted the expression of $\mathcal{L}_{GB}$ from the Einstein equation into the action. Notice that the above equation is independent of the singular coefficient $\alpha = \tilde{\alpha}/(D-4)$. Taking the limit $D\rightarrow 4$ will definitely make the action divergent, which suggests that the on-shell action is not well defined in this limit. Expectedly, the action remains well behaved for $D\geq5$.

Here one might argue that the on-shell action does usually contain divergences and the above result may not be problematic. However, the usual divergences in the action generally appear because of the infinite extent of the space time, \textrm{i.e.} $r\rightarrow \infty$, and hence those divergences are physical. The Einstein-Gauss-Bonnet theory, on the other hand, gives additional divergences in the limit $D\rightarrow 4$, which do not seem to have any physical origin.

\subsection{Surface terms}
For a well-defined variational principle, one has to supplement the action (\ref{action}) with the surface terms. These terms are required so that upon variation with metric fixed at the boundary, the action yields the Einstein equation (\ref{EOM}). Though these surface terms do not modify the field equations, however, they are essential for a well defined variational problem for a gravitational system having boundaries, like the AdS space. For the Einstein-Hilbert part, the surface term is a well known Gibbons-Hawking boundary term
\begin{eqnarray}
S_{GH}^{S.T} = -\frac{1}{8 \pi G} \int_{\partial \mathcal{M}} d^{D-1}x \ \sqrt{-\gamma} \ \mathcal{K} \,.
\label{surfaceEH}
\end{eqnarray}
here $\gamma$ is the determinate of the induced metric of the boundary $\partial \mathcal{M}$ embedded in $\mathcal{M}$ and $\mathcal{K}$ is the trace of the extrinsic curvature of the boundary. For the metric in eq.~(\ref{solg}), the Gibbons-Hawking boundary term reduces to
\begin{eqnarray}
S_{GH}^{S.T} = -\frac{\omega_{D-2}\beta}{16 \pi G} \left[ 2(D-2)r^{D-3}f+r^{D-2}f' \right]\bigg\rvert_{r\rightarrow\infty} \,.
\label{surfaceEHq}
\end{eqnarray}
where $\omega_{D-2}$ is the area of the unit $(D-2)$-dimensional sphere and $\beta$ is the inverse temperature. Similarly, the surface term counterpart of the Gauss-Bonnet Lagrangian is \cite{Myers:1987yn,Brihaye:2008xu}
\begin{eqnarray}
S_{GB}^{S.T} = -\frac{4\alpha}{16 \pi G} \int_{\partial \mathcal{M}} d^{D-1}x \ \sqrt{-\gamma} \ \left[ \mathcal{J}-2 \mathcal{G}_{ab} \mathcal{K}^{ab}  \right] \,.
\label{surfaceGB}
\end{eqnarray}
where $\mathcal{G}_{ab}$ is the Einstein tensor of boundary metric and $\mathcal{J}$ is the trace of the tensor
\begin{eqnarray}
\mathcal{J}_{ab} =\frac{1}{3} \left[2 \mathcal{K} \mathcal{K}_{ac}\mathcal{K}_{b}^{c}+\mathcal{K}_{cd}\mathcal{K}^{cd}\mathcal{K}_{ab}-2\mathcal{K}_{ac}\mathcal{K}^{cd}\mathcal{K}_{db}-\mathcal{K}^2 \mathcal{K}^{ab}  \right] \,.
\label{tensorJ}
\end{eqnarray}
Let us now explicitly evaluate $S_{GB}^{S.T}$ to see whether it is well defined in the limit $D\rightarrow 4$ or not. After a little bit of algebra, one can show that the Gauss-Bonnet surface term in $D$-dimensions simply reduces to
\begin{eqnarray}
S_{GB}^{S.T} = -\frac{\omega_{D-2}\beta}{16 \pi G} 4\alpha(D-2)(D-3) \left[- \frac{r^{D-4}f'(f-1)}{2} + (D-4) r^{D-5} f \left( 1- \frac{f}{3}  \right)   \right]\bigg\rvert_{r\rightarrow\infty} \,.
\label{surfaceGB1}
\end{eqnarray}
We see that the surface terms associated with the Gauss-Bonnet term can be divided into two parts. Those which contain a multiplicative $(D-4)$ factor and those which do not. The redefinition of the coupling constant $\alpha = \tilde{\alpha}/(D-4)$ and the subsequent limit $D\rightarrow 4$ are well defined for those terms which contain a multiplicative $(D-4)$ factor. However, the same can not be said for those terms which do not contain a multiplicative factor of $(D-4)$. Overall, like for the on-shell action, the limit $D\rightarrow4$ makes the Gauss-Bonnet surface term divergent.\\

Further, note that the surface terms for the Gauss-Bonnet action exist in all dimensions, including $D=4$. They are also finite and well-behaved for the original Gauss-Bonnet coupling (see eq.~(\ref{surfaceGB1})) and, as such, there is no problem with them in $D=4$ dimensions as well. However, the problem arises when rescaling in the coupling constant $\alpha=\tilde{\alpha}/(D-4)$ is performed. In this case, some of the surface terms give un-physical divergence in the limit $D\rightarrow4$.\\

One might wonder whether the sum of the on-shell and the surface terms can make the Gauss-Bonnet contribution to the total action finite in the limit $D\rightarrow 4$. To analyse this,  let us evaluate
\begin{eqnarray*}
& & S_{Total} =  S_{EGB}^{\text{on-shell}} + S_{GH}^{S.T} + S_{G.B}^{S.T} \, \nonumber\\
& & \ \ \ \ \ \ \  \ \ = \frac{\omega_{D-2}\beta}{16 \pi G} \left[ \int_{rh}^{\infty} \frac{2(D-1)}{L^2} r^{D-2} dr -\left( 2(D-2)r^{D-3}f+r^{D-2}f'  \right)\bigg\rvert_{r\rightarrow\infty} \right] \nonumber\\
& & \ \ \ \ \ \ \ \ \ \ - \frac{\omega_{D-2}\beta}{16 \pi G} \biggl[\int_{rh}^{\infty} \frac{2 \alpha  r^{D-2}\mathcal{L}_{GB}}{D-2} \ dr + 4\alpha(D-2)(D-3)(D-4) \times \nonumber\\
& & \hspace{5cm} \left(\frac{r^{D-4}f'(1-f)}{2(D-4)} +r^{D-5}f\left(1-\frac{f}{3}\right)\right)\bigg\rvert_{r\rightarrow\infty} \biggr]
\end{eqnarray*}
On substituting $\mathcal{L}_{GB}$ from  eq.~(\ref{LGBD}) and simplifying, we get
\begin{eqnarray}
& & S_{Total}  = \frac{\omega_{D-2}\beta}{16 \pi G} \left[-\frac{2r_{h}^{D-1}}{L^2} + \left( \frac{2r^{D-1}}{L^2} - 2(D-2)r^{D-3}f - r^{D-2}f'  \right)\bigg\rvert_{r\rightarrow\infty} \right] \nonumber\\
 & & \ \ \ \ \ \ \ \ \  - \frac{\omega_{D-2}\beta}{16 \pi G} \biggl[\int_{rh}^{\infty} 2\alpha (D-3)\left[ \frac{2\left[f'(f-1)r^{2(D-4)}\right]'}{r^{D-4}} + (D-4)(D-5)(f-1)^2 r^{D-6} \right]\ dr \nonumber\\
 & & \ \ \ \ \ \ \ \ \ \ \ +  4\alpha(D-2)(D-3)(D-4) \left(\frac{r^{D-4}f'(1-f)}{2(D-4)} +r^{D-5}f\left(1-\frac{f}{3}\right)\right)\bigg\rvert_{r\rightarrow\infty} \biggr] \nonumber\\
\label{stotal}
\end{eqnarray}
The above equation can be further simplified by evaluating the integrals. Using the integration by parts method, we get
\begin{eqnarray*}
& & \int_{rh}^{\infty} \left[ \frac{2\left[f'(f-1)r^{2(D-4)}\right]'}{r^{D-4}} + (D-4)(D-5)(f-1)^2 r^{D-6} \right] dr = \nonumber\\
& &  \bigg\rvert 2f'(f-1) r^{D-4} \bigg\rvert_{r=r_h}^{r=\infty} + (D-4)\int_{rh}^{\infty} 2f'(f-1)r^{D-5} \ dr \nonumber\\
& & + (D-4)(D-5) \biggl[\bigg\rvert \frac{(f-1)^2r^{D-5}}{D-5} \bigg\rvert_{r=r_h}^{r=\infty} -
\int_{rh}^{\infty} \frac{2f'(f-1)r^{D-5}}{D-5} \ dr \biggr]
\end{eqnarray*}
Notice that the second and fourth integral terms cancel out. Therefore,
\begin{eqnarray}
\int_{rh}^{\infty} \left[ \frac{2\left[f'(f-1)r^{2(D-4)}\right]'}{r^{D-4}} + (D-4)(D-5)(f-1)^2 r^{D-6} \right]\ dr = \nonumber\\
\bigg\rvert 2f'(f-1) r^{D-4} + (D-4) (f-1)^2 r^{D-5} \bigg\rvert_{r=r_h}^{r=\infty}
\label{intsimplfy}
\end{eqnarray}
Substituting eq.~(\ref{intsimplfy}) into eq.~(\ref{stotal}) and simplifying, we finally get
\begin{eqnarray}
& & S_{Total}  = \frac{\omega_{D-2}\beta}{16 \pi G} \left[-\frac{2r_{h}^{D-1}}{L^2} + \left( \frac{2r^{D-1}}{L^2} - 2(D-2)r^{D-3}f - r^{D-2}f'  \right)\bigg\rvert_{r\rightarrow\infty} \right] \nonumber\\
& & - \frac{\omega_{D-2}\beta}{16 \pi G} \biggl[  4\alpha (D-3) r_{h}^{D-4} f'(r_h)  - 2\alpha (D-3) (D-4) r_{h}^{D-5} + 2\alpha (D-3)(D-4) \times  \nonumber\\
& & \ \ \ \ \ \ \ \ \   \left[ (f-1)^2 r^{D-5}-f'(f-1)r^{D-4} + 2 (D-2)r^{D-5}f\left(1-\frac{f}{3}\right)  \right]\bigg\rvert_{r\rightarrow\infty} \biggr]
\label{stotal1}
\end{eqnarray}
Here we have used the fact that $f(r_h)=0$. In eq.~(\ref{stotal1}), we have rearranged Gauss-Bonnet terms in such a way that one can see the limiting behaviour clearly. In particular, there are no hidden $(D-4)$ factors in (\ref{stotal1}).  We again see that there are terms which do not contain a multiplicative $(D-4)$ factor. In particular, the first term of second line in eq.~(\ref{stotal1}). Since $f(r)$ and $f'(r)$ are well behaved functions, the redefinition $\alpha = \tilde{\alpha}/(D-4)$ and the subsequent limit $D\rightarrow 4$, therefore will again give unphysical divergences in the total action.\\

A word about the Lovelock theory in four (or lower) dimensions  is in order. Just like in the regularisation prescription of the Gauss-Bonnet term \cite{Glavan:2019inb}, one can again try to form the field equations for the higher order Lovelock terms in four dimensions. The price one has to pay for this is that the singular coupling constant has to be introduced at every order of the Lovelock theory \cite{Konoplya:2020qqh,Casalino:2020kbt}. For instance, for the Einstein-Lovelock theory of the form
\begin{eqnarray}
\mathcal{L} = -2\Lambda + \sum_{m=1}^{\bar{m}} \frac{1}{2^m}\frac{\alpha_m}{m} \delta_{\lambda_1 \sigma_1 \lambda_2 \sigma_2 \dots \lambda_m \sigma_m}^{\mu_1\nu_1\mu_2\nu_2\dots\mu_m\nu_m} R_{\mu_1\nu_1}^{\ \ \ \ \lambda_1\sigma_1}R_{\mu_2\nu_2}^{\ \ \ \ \lambda_2\sigma_2} \dots R_{\mu_m\nu_m}^{\ \ \ \ \lambda_m\sigma_m}
\end{eqnarray}
the contribution of the higher order Lovelock terms to the Euler-Lagrange equations
\begin{eqnarray}
R^\mu_\nu-\frac{R}{2}\delta^\mu_\nu-\Lambda\delta^\mu_\nu+\sum_{m=1}^{\bar{m}} \frac{1}{2^{m+1}}\frac{\alpha_m}{m} \delta_{\nu\lambda_1 \sigma_1 \lambda_2 \sigma_2 \dots \lambda_m \sigma_m}^{\mu \mu_1\nu_1\mu_2\nu_2\dots\mu_m\nu_m} R_{\mu_1\nu_1}^{\ \ \ \ \lambda_1\sigma_1}R_{\mu_2\nu_2}^{\ \ \ \ \lambda_2\sigma_2} \dots R_{\mu_m\nu_m}^{\ \ \ \ \lambda_m\sigma_m}
\end{eqnarray}
is trivially zero in four dimensions due to the antisymmetric nature of the rank $5$ Kronecker delta function. To perform the similar regularization scheme as suggested in \cite{Glavan:2019inb}, one must modify and introduce the singular coupling constants
\begin{eqnarray}
\alpha_m \rightarrow \frac{\alpha_m}{m} \frac{(D-3)!}{(D-2m-1)!}
\end{eqnarray}
to get non-zero contributions of higher order Lovelock terms in the field equations in four dimensions. However, as shown in \cite{Gurses:2020ofy} for the Gauss-Bonnet case (corresponding to $m=2$), these equations will again be ill-defined and will not have a covariant Lovelock tensor in four-dimensions. As far as the action analysis is concerned, as we have explicitly shown above for the Guess-Bonnet term, the generalised prescription of \cite{Glavan:2019inb} in higher order Lovelock theory will again give unphysical divergences in the on-shell action and surface terms in the limit $D\rightarrow4$.

Let us note that surface terms of the gravity action also induce the canonical momenta at the boundary. This canonical momenta for the AdS-Lovelock gravity is given by \cite{Miskovic:2007mg},
\begin{eqnarray}
\Pi_{i}^{j}=-\kappa \sum_{m=1}^{[(D-1)/2]} \frac{(D-2m)!m!}{2^{m+1}} \alpha_m \sum_{s=0}^{m-1} C_{s(m)} \left(\Pi_{s(m)} \right)_{i}^{j}
\end{eqnarray}
where $\kappa$ is related to the gravitational constant and
\begin{eqnarray}
\left(\Pi_{s(m)} \right)_{i}^{j} = \sqrt{-\gamma} \delta^{[j j_1 j_2...j_{2m-1}]}_{[i i_1 i_2 \dots i_{2m-1}]} \mathcal{R}^{i_1 i_2}_{j_1 j_2} \dots \mathcal{R}^{i_{2s-1} i_{2s}}_{j_{2s-1} j_{2s}}\mathcal{K}^{i_{2s+1}}_{j_{2s+1}} \dots \mathcal{K}^{i_{2m-1}}_{j_{2m-1}}
\end{eqnarray}
The coefficient $C_{s(m)}$ are
\begin{eqnarray}
C_{s(m)} = \frac{4^{m-s}}{s!(2m-2s-1)!!}
\end{eqnarray}
where the  Kronecker delta $\delta^{[j j_1 j_2...j_{2m-1}]}_{[i i_1 i_2 \dots i_{2m-1}]}$ is completely anti-symmetric in its indices. From the above equations, the contribution of Gauss-Bonnet  term ($m=2$) to the induced canonical momenta can be obtained. In terms of our original Gauss-Bonnet coupling constant $\alpha$, this is given as  \footnote{Note that in \cite{Miskovic:2007mg} an additional $(D-2m)!$ multiplicative factor is introduced in the definition of action (see Eqs.~(1.3) and (2.13) of \cite{Miskovic:2007mg}). In particular, the Gauss-Bonnet coupling constant ($\alpha$) in our work differs from the Gauss-Bonnet coupling constant ($\alpha_2$) of \cite{Miskovic:2007mg} by a factor of $(D-4)!$, \textit{i.e.}, $\alpha=\alpha_2 (D-4)!$}
\begin{eqnarray}
\Pi_{i}^{j} = - \kappa \sqrt{-\gamma}  \left[\delta_{i}^{j} \mathcal{K}-\mathcal{K}_{i}^{j} + \alpha \delta^{[j j_1 j_2 j_3]}_{[i i_1 i_2 i_3]} \left(\frac{1}{3} \mathcal{K}_{j_1}^{i_1} \mathcal{K}_{j_2}^{i_2} + \mathcal{R}^{i_1 i_2}_{j_1 j_2}  \right)\mathcal{K}_{j_3}^{i_3}  \right]
\end{eqnarray}
Notice that $\delta^{[j j_1 j_2 j_3]}_{[i i_1 i_2 i_3]}$ is zero at the boundary of four dimensional spacetime. Therefore, in four dimensions, the Gauss-Bonnet contribution to $\Pi_{i}^{j}$ identically vanishes.
Notice further that the coefficient of $\alpha \delta^{[j j_1 j_2 j_3]}_{[i i_1 i_2 i_3]}$, as it also contains various combinations of Riemann and extrinsic curvature tensors of the boundary metric, will again contain terms that do not have a multiplicative $(D-4)$ factor (just like the surface terms of the Gauss-Bonnet part of action). Therefore, it will not make any sense to rescale the coupling $\alpha=\alpha/(D-4)$ and take the limit $D \rightarrow 4$ in $\Pi_{i}^{j}$, as it will produce undefined $0/0$ expression in it. Moreover, as shown in \cite{Aoki:2020lig}, in the ADM Hamiltonian formulism of the four dimensional Gauss-Bonnet gravity, the  Weyl part of the total Hamiltonian will be undefined if the naive prescription of \cite{Glavan:2019inb} is used, \textit{i.e.} the total Hamiltonian can not be properly regularised in four-dimensional Gauss-Bonnet gravity.

This might not be very surprising. As we have shown above, the surface and counter terms are expected to diverge in the limit $D\rightarrow4$ if one uses the prescription of \cite{Glavan:2019inb}. Considering the relation between induced momenta $\Pi_{i}^{j}$ and surface/counter terms as advocated in \cite{Miskovic:2007mg}, it may not be surprising if some of these divergences also show up in $\Pi_{i}^{j}$ and in the Hamiltonian. Again, the these quantities are well-behaved for the original Gauss-Bonnet coupling and, as such, there are no problem with them in $D=4$ dimensions as well.  The problem arises when the rescaling of the coupling constant $\alpha=\alpha/(D-4)$ and the limit $D\rightarrow 4$ are performed. \\

\subsection{A word about the counter terms}
We saw above that both on-shell action and surface terms are divergent in four-dimensional Einstein-Gauss-Bonnet theory. There are mainly two different types of divergence  (i) the IR divergence because the volumes of both $\mathcal{M}$ and $\partial\mathcal{M}$ are infinite, and (ii) divergences due to the limit $D\rightarrow4$. To make sense of the total action one therefore has to regularise the action by eliminating these divergences. Remarkably, for the AdS spacetime, the IR divergences that arise in the total action are all proportional to the boundary metric. By subtracting suitable combinations of curvature scalars constructed from the boundary metric, called counterterms, one can accordingly make the total action finite. This counterterm regularization procedure has a physical interpretation in the AdS/CFT context and leads to a well-defined meaning to the notions of energy and momentum in AdS \cite{Henningson:1998gx,Balasubramanian:1999re}. For instance, the Weyl anomaly of the boundary conformal field theories can only be precisely matched with supergravity calculations by including proper counterterms in the gravity action \cite{Henningson:1998gx}. The counterterms are not only fundamental for a systematic development of the renormalized correlation functions of the boundary CFT but also of paramount importance for the reconstruction of holographic bulk spacetime from the boundary CFT data \cite{deHaro:2000vlm}. For more discussion on the importance of the counterterms in holographic renormalisation, including the Hamiltonian formalism and connection between counterterms and induced canonical momenta, see \cite{Papadimitriou:2004ap,Martelli:2002sp} (for a review, see \cite{Skenderis:2002wp}).

One therefore might try to regularise the four-dimensional Einstein-Gauss-Bonnet action by a similar counterterm procedure. However, this is not as straightforward as it seems and there are many subtleties in implementing the counterterm method. In particular, even for the Einstein action, the expression of the counterterms explicitly depend on $D$ and it changes from dimension to dimension \cite{Henningson:1998gx,Balasubramanian:1999re,deHaro:2000vlm}. The situation is even more complicated with the Gauss-bonnet action. As far as we know, the general expression of counterterms for the Gauss-bonnet action in arbitrary dimension $D$ is not known \footnote{A different counterterm regularization method, called Kounterterm regularization, for the Gauss-Bonnet gravity in $D$-dimensions  has been suggested in \cite{Olea:2006vd,Kofinas:2006hr}. In this method, the total action is rewritten as $S=S_{EGB} + c_{D-1}\int_{\partial\mathcal{M}}d^{D-1}x \ B_{D-1}$, without explicitly adding the surface terms. Here, $c_{D-1}$ is a dimension dependent constant and function $B_{D-1}$ is made up of boundary intrinsic and extrinsic curvatures. The explicit form of $B_{D-1}$, however, depends on whether $D$ is odd or even. While applying this method, our preliminary analysis suggests that depending upon whether we start from odd or even $D$, the $D\rightarrow4$ limit might not give a unique answer for the total action of four-dimensional Einstein-Gauss-Bonnet theory.  This again sounds problematic for the theory, though more work is needed for confirmation.  It will certainly be interesting to perform a detailed analysis of the total action using the Kounterterm regularization method. We leave this exercise for future work.}. There are numerous counterterms proposals to handle higher-derivative terms, but all of them seem to work in specific dimensions. Moreover, the number of terms and their complexity severally enhance with $D$ \cite{Brihaye:2008xu}. Since the whole idea of \cite{Glavan:2019inb} is based on the fact that one must first do the computation in $D$ dimensions and then take the limit $D\rightarrow4$,  therefore, to obtain consistent counterterms for the four-dimensional Einstein-Gauss-Bonnet theory one must first evaluate them in $D$ dimensions. This is an extremely non-trivial task, as infinitely many terms can contribute to the counterterms in general $D$. We can certainly evaluate the counterterms for a fixed $D$ (say $D=8$), however, then it would not make sense to let it go to four.

One might also try to find the counterterms using an ad-hoc way, for example by guessing them, such that all the divergences in the four-dimensional Einstein-Gauss-Bonnet action cancel out. This is the usual working procedure for gravity theories in AdS space. However, this ad-hoc procedure can not be called physical and considered seriously in the context of four-dimensional Einstein-Gauss-Bonnet theory, as the counterterms are then not obtained from a consistent $D\rightarrow4$ limit, \textit{i.e.} this ad-hoc procedure will be against the very philosophy of \cite{Glavan:2019inb}.

Before concluding this section, we like to stress that the counterterms are also necessary to make the variational problem well defined in AdS gravity with Dirichlet boundary conditions \cite{Papadimitriou:2005ii}, and their importance are not just limited to provide IR finiteness. In particular, since the induced metric contains a second order pole and diverges at the boundary, Dirichlet boundary condition in AdS spaces implies that a conformal class of metrics must be kept fixed at the boundary, \textit{i.e} it is only the conformal class of metrics that is well-defined at the boundary. As is well known from the holographic renormalisation procedure  \cite{Papadimitriou:2005ii}, keeping this conformal structure fixed at the boundary under a variational problem uniquely determine the nature of counterterms in Einstein gravity in general $D$-dimensions.

One may try to find the counterterms in Einstein-Gauss-Bonnet theory in a similar way. However, since the expressions of the counterterms, in general, are dimension dependent (and, moreover, they depend on whether $D$ is even or odd) from the holographic renormalisation procedure as well, one may again face ambiguity in defining them in the limit $D\rightarrow4$.  It would certainly be interesting to compute the counterterms in Gauss-Bonnet theory using the holographic renormalisation procedure and write them explicitly in terms of $D$ to see whether the limit $D\rightarrow4$ is uniquely defined or not. This will be a challenging task, considering that non-trivial anomaly terms appear only in even dimensions  (and those too vary from dimensions to dimensions) and not in odd dimensions, \textit{i.e} rearrangement of the boundary counterterms in terms of general $D$ may not be straightforward. We hope to report on these and other topics soon.

\section{Concluding remarks}
Recently, a novel Einstein-Gauss-Bonnet theory in four dimensions was suggested which not only bypasses the Lovelock's theorem but also contains the same number of massless spin-2 degrees of freedom as the Einstein-Hilbert term. This theory was defined as a limiting case of the original $D$-dimensional Einstein-Gauss-Bonnet theory with rescaled coupling constant $\alpha=\tilde{\alpha}/(D-4)$. The main idea was that the $(D-4)$ factor in the singular coefficient $\alpha$ can cancel the $(D-4)$ factor that generally appears in the Einstein equations. In this note, we further scrutinised this idea at the action level. We investigated the on-shell action and the corresponding surface terms and showed that these terms are not finite in the limit $D\rightarrow4$. In particular, the singular coefficient $\alpha$ makes the total Einstein-Gauss-Bonnet divergent in the $D\rightarrow4$ limit. The four-dimensional Einstein-Gauss-Bonnet theory therefore seems to be imprecise at least at the action level. We further highlighted various issues related to the counterterms regularisation in the four-dimensional Einstein-Gauss-Bonnet theory.

The premise of our work is based on the fact that the Einstein-Gauss-Bonnet action itself is not sufficient to be considered as a well-defined gravity theory and it has to be supplemented by the various surfaces and counterterms -- otherwise, there is no well-defined variation problem to the action itself, which essentially makes the theory ill-defined. These terms, therefore, have to be discussed thoroughly. We analyse these terms carefully in the $D$-dimensional Einstein-Gauss-Bonnet gravity and find evidence of un-physical divergences in the limit $D$ going to four, without assuming any validity of \cite{Glavan:2019inb}.

At this point, one might say that though the surface terms are essential for a well-defined variation problem, however, once the variation is done and the desired equation of motion is found, the surface terms do not play a significant part further and, therefore, may not be of much importance to gravitational theories for which the field equations are taken to be the defining object (such as the novel $4D$ Einstein-Gauss-Bonnet theory). Here, we like to strongly emphasize that the surface terms are of paramount importance to any gravitation theory and their objectives are not just restricted to provide a good boundary value variation problem. In particular, the surface terms are fundamental to the path integral formulation of quantum gravity \cite{Hawking:1978jz}. The Hamiltonian formulation of gravity theory further necessitates the need for the surface terms \cite{York:1972sj}. Indeed, there are many reasons to consider the action (with appropriate surface terms included), as the more fundamental object in gravitational theory. What our analysis suggests here is that it raises many difficulties in interpreting the resultant $D\rightarrow4$ limiting action (of the original $D$-dimensional Einstein-Gauss-Bonnet action) as the defining action for the $4D$ Einstein-Gauss-Bonnet field solutions, since the action itself is ill-defined in  this limit. This result also complements the recent findings \cite{Fernandes:2020nbq,Hennigar:2020lsl,Aoki:2020lig,Alkac:2020zhg,Casalino:2020pyv,Bonifacio:2020vbk,Arrechea:2020evj,Liu:2020evp,Lu:2020iav}, which do suggest a different action, in particular the scalar-tensor type action, for the $4D$ Gauss-Bonnet solution.

Another issue that deserves further attention is the requirement of the finiteness of the on-shell action, especially for those gravitation theories which are formulated at the equation of motion level. In particular, if all the necessary results are derivable from the field equations alone then the ill-defined nature of the on-shell action might not seem problematic. To some extent, this argument may sound legit at-least at the classical level, where important objects like the conserved quantities can be constructed from the field equations. However, there are many reasons to believe that the scope of complete understanding of the gravity theory using the field equations alone is severely limited and they need be supplemented by a proper action for a concrete description. In particular, eventually, we would like to quantise the gravity theory and study its spectrum. There, it would not be possible to make much progress from the field equations alone. Indeed, the four-dimensional Einstein-Gauss-Bonnet theory was suggested as a classical theory of gravity. If this novel four-dimensional gravity theory turns out to be a genuine alternative to the Einstein gravity (though its field equations are itself questionable \cite{Gurses:2020ofy}, and therefore is highly unlikely), then it would certainly be desirable to have its quantum spectrum, which may not be possible to compute just from its field equation.

Further, the on-shell action also appears naturally in the path integral and hamiltonian formulation of quantum gravity. In particular, in the path integral formulation, one would expect that the dominant contribution in the partition function (and other transition amplitudes) would come from the extremum action \textit{i.e.} the action obtained from the solutions of the field equations. From the partition function (via the on-shell action), one can then directly compute the spectrum, conserved quantities and other important objects of theory from first principle. In fact, the same methodology is used in computing important observables in some of the most trustworthy quantum gravity models, for instance in the AdS/CFT correspondence \cite{Maldacena:1997re}. Indeed, the dual boundary CFT information in the large $N$ limit ($N$ being the number of colours) is generally encoded in the gravity on-shell action in the AdS/CFT framework \cite{Witten:1998qj,Gubser:1998bc}.

An interesting question one might ask is, does the novel four-dimensional Einstein-Gauss-Bonnet theory in AdS space exhibit a dual boundary theory. In the $AdS_D/CFT_{D-1}$ context, the gravity and the dual boundary theory are connected in the semiclassical approximation via
\begin{eqnarray}
& & Z_{CFT}=Z_{AdS}= e^{-S_{AdS}}\,.
\end{eqnarray}
where $e^{-S_{AdS}}$  is the classical gravitational action. In this approximation, the AdS action becomes the generating function of the connected correlation functions of dual CFT. Since the gravity action is directly related to the physical observables of the dual CFT theory, it is desirable that $S_{AdS}$ remains free from any divergences. Note that the usual IR divergences in the gravity side correspond to UV divergences in the dual CFT side and therefore have a precise meaning \cite{deHaro:2000vlm}. However, the same can not be said for the divergences that appear due to $D\rightarrow4$ limit. In particular, the introduction of Gauss-Bonnet term in the gravity action corresponds to next to the leading order corrections to the $1/N$ expansion of the dual CFT \cite{Nojiri:2000gv}. Therefore, the four-dimensional Einstein-Gauss-Bonnet action should not contain any divergence whose dual counterpart in $CFT_3$ does not exist. As we have shown in this work, unless the counterterms miraculously cancel out $D\rightarrow4$ divergences, it seems difficult to make a dual CFT connection of the novel four-dimensional Einstein-Gauss-Bonnet theory.\\

\section*{Acknowledgments}
We would like to thank B.~Tekin for useful discussions. We would like to thank D.~Choudhuri for careful reading of the manuscript and pointing out the necessary corrections. The work of SM is supported by the Department of Science and Technology, Government of India under the Grant Agreement number IFA17-PH207 (INSPIRE Faculty Award).

\end{document}